\documentclass[a4paper,twocolumn,11pt]{quantumarticle}
\pdfoutput=1
\usepackage[utf8]{inputenc}
\usepackage[english]{babel}
\usepackage[T1]{fontenc}
\usepackage{amsmath}
\usepackage{hyperref}
\usepackage[numbers,compress]{natbib}
\usepackage{tikz}
\usepackage{lipsum}

\usepackage{graphicx}
\usepackage{float}
\usepackage{dcolumn}
\usepackage{bm}
\usepackage{amsfonts,amssymb,amscd,amsmath,amsthm}
\usepackage{algorithm}
\usepackage{algpseudocode}
\usepackage{enumerate}
\usepackage{epsfig}
\usepackage{subfigure}
\usepackage{physics}
\usepackage{svg}
\usepackage{tikz}
\usepackage[colorlinks = true]{hyperref}
\hypersetup{colorlinks=true, linkcolor=blue, citecolor=blue}
\usepackage{amsmath, amsthm, amsfonts, amssymb, amsbsy, mathtools, xcolor, bm, bbm, graphicx, changepage, cleveref}
\usepackage{xcolor}
\usepackage{physics}
\usepackage{epstopdf}
\usepackage{framed}
\usepackage{multirow}
\usepackage{color}
\usepackage{longtable}
\usepackage{comment}
\usepackage{qcircuit}
\usepackage{faktor}

\begin{document}

\title{Gaussian boson sampling validation via detector binning}

\author{Gabriele Bressanini}
\email{g.bressanini@imperial.ac.uk}
\affiliation{Blackett Laboratory, Imperial College London, London SW7 2AZ, United Kingdom}

\author{Benoit Seron}
\email{benoitseron@gmail.com}
\affiliation{Quantum Information and Communication, Ecole polytechnique de Bruxelles, CP 165/59, Université libre de Bruxelles (ULB),
1050 Brussels, Belgium}
\affiliation{Physikalisches Institut, Albert-Ludwigs-Universit\"at Freiburg,
Hermann-Herder-Stra{\ss}e 3, D-79104 Freiburg, Germany}
\affiliation{EUCOR Centre for Quantum Science and Quantum Computing,
Albert-Ludwigs-Universit\"at Freiburg, Hermann-Herder-Stra{\ss}e 3, D-79104 Freiburg, Germany}

\author{Leonardo Novo}
\email{lfgnovo@gmail.com}
\affiliation{International Iberian Nanotechnology Laboratory (INL), Av. Mestre José Veiga, 4715-330 Braga, Portugal}

\author{Nicolas J. Cerf}
\email{ncerf@ulb.be}
\affiliation{Quantum Information and Communication, Ecole polytechnique de Bruxelles, CP 165/59, Université libre de Bruxelles (ULB),
1050 Brussels, Belgium}

\author{M.S. Kim}
\affiliation{Blackett Laboratory, Imperial College London, London SW7 2AZ, United Kingdom}

\begin{abstract}
Gaussian boson sampling (GBS), a computational problem conjectured to be hard to simulate on a classical machine, has been at the forefront of recent years' experimental and theoretical efforts to demonstrate quantum advantage. The classical intractability of the sampling task makes validating these experiments a challenging and essential undertaking.
In this paper, we propose binned-detector probability distributions as a suitable quantity to statistically validate GBS experiments employing photon-number-resolving detectors.
We show how to compute such distributions by leveraging their connection with their respective characteristic function.
The latter may be efficiently and analytically computed for squeezed input states as well as for relevant classical hypothesis like squashed states.
Our scheme encompasses other validation methods based on marginal distributions and correlation functions.
Additionally, it can accommodate various sources of noise, such as losses and partial distinguishability, a feature that has received limited attention within the GBS framework so far.
We also illustrate how binned-detector probability distributions behave when Haar-averaged over all possible interferometric networks, extending known results for Fock boson sampling.  
\end{abstract}

\maketitle

\section{Introduction}

Gaussian boson sampling (GBS) \cite{GBS_original} is a well defined computational problem  that, under plausible complexity-theoretic 
assumptions, is conjectured to be hard to simulate (even approximately) by classical means \cite{Grier2022complexityof,GBScomplexity}.
The task consists of sampling from the output state of a passive linear optical network (LON) fed with squeezed light, using photon-number-resolving (PNR) detectors.
Recent progress in the field of photonic quantum technology led to multiple independent claims of quantum advantage  \cite{zhong2020science, zhong2021phase, madsen2022xanadu, deng2023gaussian}. 
In addition to constituting a prime candidate for an experimental demonstration of quantum advantage using present day technological capabilities, GBS also finds application in solving problems of practical interest such as simulating molecular vibronic spectra \cite{vibronic, oh2022quantum}, predicting stable molecular docking configurations for drug development \cite{molecular_docking}, perfect matchings counting \cite{perfect_matchings_counting} and finding dense subgraphs \cite{ArrazolaDense}.

A fundamental problem of GBS is that of verifying the correct functioning of the device.
That means, we want to certify that data samples are drawn from the ideal theoretical distribution (also known as \textit{ground truth}), and not from an efficiently computable distribution that only resembles it.
For small systems, the ground truth can be analytically computed for arbitrary input Gaussian states and may therefore be directly compared with the experimental observations \cite{zhong2020science}.
However, despite remarkable progress of classical algorithms for the simulation of a boson sampler \cite{bulmer2022boundary, quesada2022quadratic, quesada2020exact, gupt2020classical, bourassa2021fast, chabaud2023resources, oh2022classical}, in the quantum advantage regime direct comparison with the ground truth is hindered by the very nature of the problem.
In fact, computing the theoretical distribution involves the evaluation of the Hafnian of complex matrices, a problem known to be \#P-hard.
Additionally, even if we had access to the ground truth, the problem would persist, as an exponential number of samples would be needed to experimentally estimate the output probabilities. For these reasons, full certification is believed to be out of reach \cite{hangleiter_sample_complexity} and 
one has to rely on indirect methods to probe the correct functioning of a Gaussian boson sampler. 
In particular, validation protocols based on the evaluation of an efficiently computable quantity aim at identifying scalable and efficient statistical tests that any GBS experiment operating in the quantum advantage regime is expected to pass.
Useful validation methods should have the following desirable properties \cite{walschaers2020signatures, flamini2020validating}. 
First, they should be \emph{universal} $-$ i.e. applicable to any interferometric setup $-$ an especially important requirement for GBS applications that need configurability of the LON.
They should also be sensitive to high-order multi-photon interference \cite{shchesnovich2022boson}, an effect that can be hampered by partial distinguishability of the input states \cite{renema2020simulability, Shchesnovich_2022_partial_dist_gaussian, shi2022effect}.
Finally, any practical validation protocol must make a limited use of resources.
In particular, the protocol must be \emph{computationally efficient}, meaning that the quantity at its core only requires polynomially many calculations for its evaluation on a classical computer, and must be \emph{sample efficient}, i.e. it requires polynomially many experimental data samples to estimate the quantity to meaningful relative precision.
\begin{figure}[t]
    \centering
    \includegraphics[width = 0.5\textwidth]{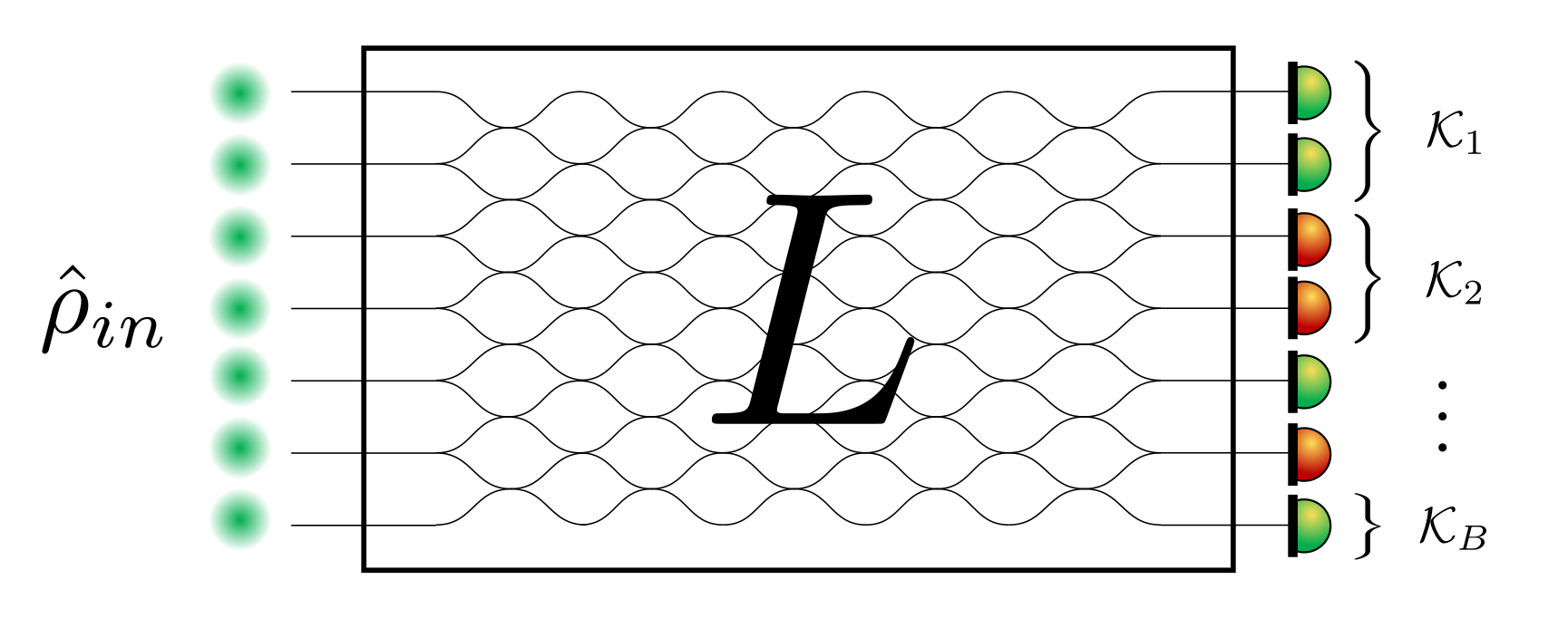}
    \caption{\textbf{Binned-detectors GBS.} A multi-mode Gaussian input state $\hat{\rho}_{in}$ is sent through a linear optical network described by the matrix $L$. The output state's photon number distribution is sampled from using PNR detectors. These are grouped together into $B$ bins according to the partition $\lbrace \mathcal{K}_j\rbrace_{j=1}^B$ of the output modes.  $P(k_1,\dots,k_B)$ denotes the binned probability distribution, i.e. the probability of measuring $k_j$ photons in the $j-$th bin.}
    \label{fig:binning}
\end{figure}

Several validation methods for GBS are found in the literature. 
As already mentioned, for systems of modest size one is able to compute the full theoretical distribution and directly statistically compare it with data coming from the experiment.
For intermediate size GBS setups, Bayesian techniques \cite{flamini2020validating} may be employed.
These methods involve computing the probability of obtaining a specific output detection pattern under different hypotheses (i.e. different initial states and/or noise models), and with just a few dozen samples it becomes feasible to select the most likely hypothesis with high degree of confidence.
While these methods provide strengthening evidence for reduced versions of a larger experiment, they become highly unscalable as we approach the quantum advantage regime.
For systems operating in the quantum advantage regime, a popular choice is to consider validation methods based on the computation of low-order correlation functions between the photons counts in each output mode \cite{phillips2019benchmarking}. 
These techniques, while practical and efficient, were shown to be insufficiently sensitive to high-order multi-photon interference, as these correlators mat easily be reproduced by classical models \cite{villalonga2021efficient}.

Additional noteworthy validation methods include heavy outcome generation tests and other cross entropy benchmarks \cite{hangleiter2023computational, oh2023spoofing}, as well as that presented in Ref.~\cite{giordani2022certification}, where the connection between graph theory and GBS is exploited to verify the correct functioning of the device.

Recently, validation methods based on detector binning have been proposed and successfully applied to Fock state boson sampling (AABS) \cite{shchesnovich2016universality,shchesnovich2021distinguishing,seron2022validation} and to GBS experiments utilizing threshold detectors \cite{opanchuk2018simulating, drummond2022simulating, dellios2022validation}.
These validation protocols consist of grouping the detectors at the LON's output into a few bins (whose number must not scale with the size of the system), their measurement readings summed into a single count for each bin.
One is then interested in the probability distribution of such binned counts, a quantity which is sensitive to high-order interference.
This coarse grain operation has the effect of exponentially reducing the sample space size, meaning that the binned-mode photon number probability distribution can be estimated experimentally, up to a target total variation distance, using a number of samples which scales only polynomially with the system’s size, thus ensuring sample efficiency. 
Then, crucially, for each validation method based on this paradigm to be practical, one  must show that the theoretical binned probability distribution can be computed efficiently, i.e. in polynomial time.
This framework encompasses validation methods based on correlation functions as well as marginal distributions.

Previous GBS validation protocols based on binned-count probability distributions focused on implementations of the task that employ on/off detectors and made use of numerical Monte Carlo sampling techniques \cite{drummond2022simulating,dellios2022validation}.
GBS experiments that employ PNR detectors have become increasingly popular in recent years owing to the necessity of entering higher energy regimes (with detection events with large total photon number) to achieve quantum advantage.
Additionally, PNR-capabilities are required for most of the real-world applications of GBS, such as the simulation of molecular vibronic spectra.
In this paper, we address the problem of validating a GBS experiment employing PNR detectors by developing a framework that enables the computation of binned count probability distributions for various instances of such tasks.
In particular we show that the so called characteristic function, i.e. the quantity at the core of this work, may be efficiently and analytically evaluated, making our approach both precise and easily implementable.
Within our formalism, the only free parameter and sole source of error is the energy cutoff one has to introduce due to the Gaussian nature of the initial state.
Our method can easily accommodate for losses in the LON, detection imperfections and  partial distinguishability, a source of noise well studied in AABS, but whose role has so far been given little attention in the GBS framework.
This is crucial, as experimental implementations of GBS are unavoidably affected by different sources of noise that may challenge the sampling task from entering the regime where quantum advantage is achievable. Indeed, if enough noise is present then the sampling task becomes efficiently simulable using classical algorithms \cite{rahimikeshari,classical_simulation_GBS_Quesada,bressanini2023thermalgbs}. 

An adaptation of the validation technique based on detector binning to Gaussian boson samplers was also considered in Ref.~\cite{singh2023proofofwork}, motivated by the development of a quantum Proof-of-Work (PoW) scheme for blockchain consensus. 
Our work goes beyond this adaptation by discussing in more detail the computational complexity of the method, as well as by demonstrating how to compute the binned probability distributions under different noise models and for classical mock-up distributions. 

This paper is structured as follows.
In Section \ref{sec:binned_probability}, we introduce notation, define the binned-count probability distribution and highlight its connection to the characteristic function, the quantity at the core of this paper. 
In Section \ref{sec_characteristic_GBS}, we explicitly compute the binned-count probability distribution of a GBS instance employing PNR detectors, and provide evidence that it can be done efficiently on a classical computer.
In Section \ref{sec:asymptotic} we derive the Haar-averaged asymptotic behaviour of these distributions, while in Section \ref{sec:mock_up} we show how our formalism can readily be adapted to compute the binned probability distributions for classical input states such as thermal states and squashed states. 
In Section \ref{sec:partial_distinguishability} we show how a model of partial distinguishability may be introduced into the binned GBS framework.
Lastly, in Sec.~\ref{sec:conclusions} we draw conclusions and give some final remarks.

\section{Binned probability and characteristic function}
\label{sec:binned_probability}
Let us consider a generic sampling experiment where an $m-$mode quantum state $\hat{\rho}$ is measured via PNR detection. 
As previously mentioned, the validation scheme we propose relies on grouping the detectors into bins.
More formally, we consider a partition of the $m$ output modes into $B$ bins, i.e. non-empty, mutually disjoint subsets $\lbrace\mathcal{K}_j\rbrace_{j=1}^B$ such that $\mathcal{K}_j\subset\lbrace 1,\dots,m\rbrace$. We limit the number of bins to be independent of the size of the experiment, i.e. $B = O(1)$.
We point out that this requirement is necessary to ensure computational and sample efficiency, as it will soon be clear, however the formalism and the equations presented in this paper remain valid for any partition and number of bins.
We are interested in computing the probability $P(\bm{k})$ of observing a given detection pattern $\bm{k} = (k_1,\dots,k_B)$ of the binned detectors, where $k_j$ denotes the number of photons measured in the $j-$th bin $\mathcal{K}_j$. 
To do so, one introduces the \emph{characteristic function}\footnote{Note this is not the characteristic function typically considered in the continuous variables literature, i.e. the Fourier transform of the Wigner function.},
defined via the following expectation value
\begin{equation}
    X(\bm{\eta}) = \Tr{\hat{\rho} \, e^{i\bm{\eta}\cdot\hat{\bm{N}}}} \, .
    \label{characteristic_function_definition_2}
\end{equation}
Here $\hat{\bm{N}}$ is a vector of operators defined such that 
\begin{equation}
    \bm{\eta}\cdot\hat{\bm{N}} = \sum_{j=1}^B \eta_j\hat{N}_j = \sum_{j=1}^B \eta_ j\sum_{\ell\in\mathcal{K}_j}\hat{n}_\ell \, ,
\end{equation}
where $\hat{n}_j = \hat{a}^\dagger_j \hat{a}_j$ is the bosonic number operator of mode $j$.
We anticipate that, if $\hat{\rho}$ is a Gaussian state, then Eq.~\eqref{characteristic_function_definition_2} may be computed exactly.
In Appendix \ref{appendix_characteristic_expectation} we show how the characteristic function and the binned probability distribution are related via a discrete Fourier transform, namely
\begin{align}
    X(\bm{\eta}) &= \sum_{\bm{k}}P(\bm{k})e^{i\bm{\eta}\cdot\bm{k}}\\
    &=\sum_{\bm{k}\in\Omega^B}P(\bm{k})e^{i\bm{\eta}\cdot\bm{k}}+ \epsilon(\bm{\eta}) \,,
    \label{characteristic_function_definition_1}
\end{align}
where $\Omega^B = \lbrace \bm{k} \vert k_i \in \lbrace 0,\dots,n\rbrace,\forall i\in\lbrace 1,\dots,B\rbrace\rbrace$.
Here, $n$ is a photon-number (energy) cutoff the needs to be introduced due to the Gaussian nature of the input state which implies that the total number of photons is not fixed. The restriction to a multidimensional Fourier transform over $(n+1)^B$ points lead to an error $\epsilon(\bm{\eta})$ that decreases exponentially with $n$. As shown in Appendix \ref{appendix:cutoff}, we can choose the cutoff as 
\begin{equation}
    n= m \sinh^2(r) + 4 \cosh^2{r}\log\left(\frac{1}{\epsilon}\right)
    \label{ncutoff}
\end{equation}
to ensure that the error is bounded as follows:
\begin{equation}
   |\epsilon(\bm{\eta})|\leq \mathbb{P}\left[k>n\right] \leq \epsilon. 
   \end{equation}
Here, $\mathbb{P}\left[k>n\right]$ denotes the probability of having more than $n$ photons at the input of the boson sampler. 
From Eq.~\eqref{characteristic_function_definition_1}, an approximation of the probability $P(\bm{k})$ can then be  retrieved by means of inverse discrete Fourier transform, namely
\begin{equation}
    \tilde{P}(\bm{k}) =\frac{1}{(n+1)^B}\sum_{\bm{\nu}\in\Omega^B} X(\tfrac{2\pi}{n+1}\bm{\nu}) e^{-\frac{2\pi i}{n+1}\bm{\nu}\cdot\bm{k}}+ \tilde{\epsilon}(\bm{k})
    \label{prob_pnr}
\end{equation}
where the error $\tilde{\epsilon}(\bm{k})$ is also bounded by $\epsilon$ for the choice of the cutoff from Eq.~\eqref{ncutoff}.  

Marginal distributions are naturally encompassed by this formalism. In particular, $\ell-$marginals involve considering only a fixed set of $\ell$ output modes, while disregarding the rest (one is typically interested in single and two-mode marginals). These can be regarded as specific instances of binned probability distributions, where we consider $\ell < m$ bins comprising of a single detector each. As an example, focusing on the first $\ell$ output modes, the relevant characteristic function is
\begin{equation}
    X(\bm{\eta}) = \Tr_\ell\lbrace\Tr_{m-\ell}\lbrace\hat{\rho}\rbrace \,e^{i\sum_{j=1}^\ell \eta_j \hat{n}_j}\rbrace \, , 
\end{equation}
where $\Tr_{m-\ell}\lbrace\hat{\rho}\rbrace$ is the reduced $\ell-$mode state.

Analogously, if we instead consider threshold photo-detection, the formalism still stands, provided that we replace the operator $\bm{\eta}\cdot\hat{\bm{N}}$ in the characteristic function definition Eq.~\eqref{characteristic_function_definition_2} with 
\begin{equation}
    \bm{\eta}\cdot \hat{\bm{\Pi}} = \sum_{j=1}^B \eta_ j\sum_{\ell\in\mathcal{K}_j}\hat{\Pi}_ {1,\ell} \, ,
\end{equation}
where $\hat{\Pi}_{1,\ell} = \mathcal{I}-\ketbra{0}$ represents the "on" element of the threshold detection's POVM acting on the Hilbert space of mode $\ell$ and $\mathcal{I}$ denotes the identity operator.
Notice how, in this scenario, there is no need to introduce an energy cutoff, as the sample space size is naturally finite, regardless of the state $\hat{\rho}$. However, as opposed to GBS with PNR detection, in this case we are not able to analytically compute the characteristic function. Nevertheless, this problem may be tackled using the Monte Carlo techniques developed in Ref.~\cite{drummond2022simulating}. 
We also note that, for GBS experiments employing threshold detectors and operating in the non-collisional regime (i.e. the probability of observing two or more photons in any given output mode is negligible), we may still approximate the binned probability distribution using Eq.~\eqref{prob_pnr}.

In the next section we show that it is possible to analytically compute the characteristic function Eq.~\eqref{characteristic_function_definition_2} of a GBS experiment, by exploiting the phase-space formulation of quantum optics.

\section{Characteristic function of GBS}
\label{sec_characteristic_GBS}
A GBS experiment consists of injecting squeezed vacuum states into a passive LON, and sampling the output state using PNR detectors. 
The $m-$mode initial state $\hat{\rho}_{in}$ entering the interferometer thus reads
\begin{equation}
    \hat{\rho}_{in} =  \bigotimes_{j=1}^m \hat{S}(r_j)\ketbra{0}\hat{S}^\dagger (r_j) \, ,
    \label{initial_state}
\end{equation}
where 
\begin{equation}
    \hat{S}(r_j) = e^{{\frac{r_j}{2}(\hat{a}_j^{\dagger 2}-\hat{a}_j^2)}}
\end{equation}
is the well-known single-mode squeezing operator and $r_j>0$ is the squeezing parameter.
The evolution of the input state through the LON is described by the quantum CP-map $\mathcal{E}$.
In particular, the linear transformation of the system's modes induced by the (possibly lossy) LON is entirely characterized by a sub-unitary matrix $L$. 
In the following, we provide an overview of the techniques employed to analytically compute the characteristic function  
\begin{equation}
    X(\bm{\eta}) = \Tr{\mathcal{E}(\hat{\rho}_{in}) e^{i\bm{\eta}\cdot\hat{\bm{N}}}} \, ,
\end{equation}
while the details of the calculation may be found in Appendix \ref{appendix_detailed_calculation}.

Using the identity $e^{i\theta\hat{n}} = :e^{(e^{i\theta}-1)\hat{n}}:$, where $:\bullet:$ denotes normal operator ordering, we can write the multi-mode phase-shift operator $e^{i\bm{\eta}\cdot\hat{\bm{N}}}$ as follows
\begin{equation}
    e^{i\bm{\eta}\cdot\hat{\bm{N}}} = \bigotimes_{j=1}^B \bigotimes_{\ell\in\mathcal{K}_j} :e^{(e^{i\eta_j}-1)\hat{n}_\ell} : \, .
    \label{phase_shift1}
\end{equation}
The expectation value of a normally-ordered operator may then be evaluated by averaging over the phase-space variables according to the positive $P$ representation of the state.
In fact, any $m-$mode quantum state $\hat{\rho}$ admits a non-negative phase-space representation via a quasi-probability distribution $P(\bm{\alpha},\bm{\beta})$ such that
\begin{equation}
    \hat{\rho} = \int_{{\mathbb{C}^{2m}}} d^{2m}\bm{\alpha}\,d^{2m}\bm{\beta} P(\bm{\alpha,\bm{\beta}})\, \frac{\ket{\bm{\alpha}}\!\!\bra{\bm{\beta}^*}}{\braket{\bm{\beta}^*\vert{\bm{\alpha}}}} \, ,
\end{equation}
where $\ket{\bm{\alpha}} = \ket{\alpha_1}\otimes \cdots \otimes \ket{\alpha_m}$ is an $m-$mode coherent state. 
Furthermore, a squeezed vacuum state admits a positive $P$ representation on the real space (rather then complex) \cite{drummond2022simulating}, and consequently the positive $P$ representation of $\hat{\rho}_{in}$ reads 
\begin{equation}
    P_{in}(\bm{x},\bm{y}) = \prod_{i=1}^m \left[ \frac{\sqrt{1+\gamma_i}}{\pi\gamma_i} e^{-(x_i^2+y_i^2)(\gamma_i^{-1}+{1}/{2})+x_i y_i} \right] ,
\end{equation}
where $\bm{x},\bm{y}\in\mathbb{R}^m$ and $\gamma_i = e^{2r_i} - 1$.
After some calculations, we obtain 
\begin{equation}
\begin{split}
    X(\bm{\eta})  = 
    \int_{{\mathbb{R}^{2m}}} & d^m\bm{x} \, d^m\bm{y} P_{in}(\bm{x},\bm{y}) \\ & 
     e^{\sum_{j=1}^B (e^{i\eta_j}-1) \sum_{\ell\in\mathcal{K}_j}(L^*\bm{y})_\ell(L\bm{x})_\ell} \, .
\end{split}
\end{equation}
Standard, multi-dimensional Gaussian integration yields the final result
\begin{equation}
\label{eq:X_eta_determinant}
   X(\bm{\eta}) = \prod_{i=1}^m \left[ \frac{2\sqrt{1+\gamma_i}}{\gamma_i}\right]\frac{1}{\sqrt{\det{Q}}}\, ,
\end{equation}
where the matrix $Q$ is defined as 
\begin{equation}
    Q  = 
    \begin{pmatrix}
        2\Gamma^{-1} + \mathbb{I}_m  & -L^\intercal \text{diag}\lbrace e^{i\theta_j} \rbrace_{j=1}^m L^* \\
        -L^\dagger \text{diag}\lbrace e^{i\theta_j} \rbrace_{j=1}^m L &  2\Gamma^{-1} + \mathbb{I}_m 
        \end{pmatrix} \, , 
    \, 
\end{equation}
with $\Gamma = \text{diag}\lbrace \gamma_i \rbrace_{i=1}^m$.

Let us now briefly consider the computational complexity of calculating the  binned probability distribution Eq.~\eqref{prob_pnr} using the approach outlined above.
As previously pointed out, Gaussian states do not have a definite photon content, hence a suitable energy cutoff $n$ needs to be introduced to guarantee a bounded error (see Eq.~\eqref{ncutoff}). 
With this constraint, only $(n+1)^B$ binned photon-count patterns are taken into account, and from Eq. \eqref{prob_pnr} we see that we need to evaluate the characteristic function in $(n+1)^B$ points in order to approximate the binned-count probability distribution.
Lastly, Eq.~\eqref{eq:X_eta_determinant} reveals that computing $X(\bm{\eta})$ amounts to evaluating the determinant of the $2m * 2m$ matrix $Q$, which can be done exactly in a time scaling polynomially with $m$.
From these considerations, we conclude that the calculation of binned probability distributions is computationally efficient. In fact, the dependency of the complexity on the target error $\epsilon$ is exponentially better ($\text{poly}{(\log{(1/\epsilon)}})$) when compared to the counterpart of this validation method for standard boson sampling ($\text{poly}{(1/\epsilon)}$) \cite{seron2022validation}. In the latter case, even though there is no need for an energy cutoff as the total number of photons is fixed, the characteristic function is given by a permanent and so its approximation up to error $\epsilon$ takes $\text{poly}{(1/\epsilon)}$ time.

\section{Haar-averaged distributions}
\label{sec:asymptotic}
Within the usual paradigmatic setting of ideal GBS, the $m\times m$ unitary matrix $U$ that describes the LON is drawn at random according to the Haar measure. In this section we derive the asymptotic properties of the binned probability distribution $P(\bm{k}\vert U)$ averaged over all possible interferometric configurations, where we have highlighted the $U$-dependence of the function. We also derive the corresponding distribution for distinguishable input states.

We remind the reader that the Haar-average of $P(\bm{k}\vert U)$ is defined as 
\begin{equation}
    \expval{P(\bm{k}\vert U)} = \int P(\bm{k}\vert U) d\mu(U) \, ,
\end{equation}
where $d\mu$ denotes the Haar measure and the integral is taken over the whole unitary group. 
The unitary invariance of the Haar measure implies that for every unitary matrix $W$ it holds that
\begin{equation}
\begin{split}
    \expval{P(\bm{k}\vert U)} &  = \int P(\bm{k}\vert UW) d\mu(U)  \\ & = \int P(\bm{k}\vert UW) d\mu(U) d\mu(W) \, .
    \label{haar_average1}
\end{split}
\end{equation}
The second equality follows from the fact that the averaged binned probability distribution is independent of $W$, thus justifying a further average over the latter.
In what follows, we take $W$ as a diagonal matrix that represents an $m$-mode phase shift $e^{i\bm{\phi}\cdot\hat{\bm{n}}}$ applied to the input state.
Consequently, the Haar measure simply reads $d\mu(W)=d^m\bm{\phi}/(2\pi)^m$. 
It is well-known that averaging over $\bm{\phi}$ causes the off-diagonal elements of the initial state's density matrix $-$ expressed in the Fock basis $-$ to vanish.
In fact, let us consider a generic $m$-mode quantum state 
\begin{equation}
    \ket{\psi} = \sum_{\bm{n}} c_{\bm{n}} \ket{\bm{n}} \, ,
\end{equation}
where $\ket{\bm{n}} = \ket{n_1}\otimes\cdots\otimes\ket{n_m}$ denotes an $m-$mode Fock state.
One easily shows that 
\begin{equation}
\begin{split}
     & \int \frac{d^m\bm{\phi}}{(2\pi)^m}e^{i\bm{\phi}\cdot\hat{\bm{n}}}\ketbra{\psi}e^{-i\bm{\phi}\cdot\hat{\bm{n}}} 
      = \sum_{\bm{n}} \vert c_{\bm{n}} \vert^2 \ketbra{\bm{n}},
      \label{fully_decohered}
\end{split}
\end{equation}
where we have used the integral representation of the Kroneker delta 
\begin{equation}
    \int \frac{d^m\bm{\phi}}{(2\pi)^m} e^{i(\bm{n}-\bm{m})\cdot\bm{\phi}} = \delta_{\bm{n},\bm{m}} \, .
\end{equation}
Hence, the above argument implies that, when computing Haar averages of the binned probability distribution, we can equivalently substitute any initial state  with its fully decohered version.
The advantage of dealing with such statistical mixture lies in the fact that one can readily exploit known results applicable to Fock states input, by means of post-selecting on the total number of detected photons $n$.

In Ref.~\cite{shchesnovich2017asymptotic} Shchesnovich used combinatorial arguments to  prove that, given $n$ input photons $-$ either perfectly distinguishable or indistinguishable $-$ impinging on an $m$-mode unitary LON whose output modes are partitioned into $B$ bins, the probabilities of observing a specific detection pattern $\bm{k}=(k_1,\dots,k_B)$, when averaged over the Haar-random interferometers, are given by 
\begin{align}
    \label{eq:asymptoticsD}
    \expval{P^{\text{dist}}_{\text{Fock}}(\bm{k})} &= \frac{n!}{\prod_{i=1}^{B} k_i!}\prod_{i=1}^{B} q_i^{k_i} \, ,  \\
    \label{eq:asymptoticsB}
    \expval{P^{\text{indist}}_{\text{Fock}}(\bm{k})} & =\expval{P^{\text{dist}}_{\text{Fock}}(\bm{k})} \frac{\prod_{i=1}^{B} (\prod_{\ell=0}^{k_i-1} [1+\ell/\mathcal{K}_i]
    )}{\prod_{\ell=0}^{n-1} [1+\ell/m]} \, ,
\end{align}
where $q_i = \vert \mathcal{K}_i\vert/m$, $\vert \mathcal{K}_i \vert$ being the cardinality of $\mathcal{K}_i$, i.e. the number of output modes within the $i-$th bin. 
By taking the asymptotic limit $n\gg 1$, $B\ll n$ and $B\ll \min{\mathcal{K}_i}$, these expressions may be further reduced to the following Gaussian form 
\begin{align}
\label{gaussian_law}
    \expval{P^{\sigma}_{\text{Fock}}(\bm{k})} &\approx \frac{\exp \{ 
    -n \sum_{i = 1} ^B
    \frac{(x_i-q_i)^2}{2(1+\sigma \alpha)q_i}
    \}}{(2\pi(1+\sigma \alpha)n)^{(B-1)/2} \prod_{i = 1} ^B \sqrt{q_i}} \nonumber \\
    &\times \left(
    1 + \mathcal{O}
    \left( \frac{\alpha \delta_{\sigma, 1}}{n}
    \right)
    \right),
\end{align}
where $\alpha = n/m$ is the particle density, $\delta_{\sigma,1}$ denotes the the Kronecker delta, and we have $\sigma = 1$ for indistinguishable particles and $\sigma = 0$ for distinguishable ones.
Eq.~\eqref{gaussian_law} represents 
the quantum generalization of the well-known asymptotic law for a multinomial distribution (de Moivre-Lagrange-Laplace theorem \cite{hald2008history}) which governs the behaviour of boson sampling instances with perfectly distinguishable particles ($\sigma = 0$). 
Quantum statistical effects are taken into account by the parameter $\alpha$.
Note how the asymptotic law in Eq.~\eqref{gaussian_law} only depends on the total number of photons, but not on the the specific Fock input state compatible with the total photon number $n$.
Additionally, Eq.~\eqref{gaussian_law} is valid for any assignment of modes to bins, at fixed cardinality of the latter.

\begin{figure*}[t]

  \centering
    \includegraphics[width=0.95\textwidth]{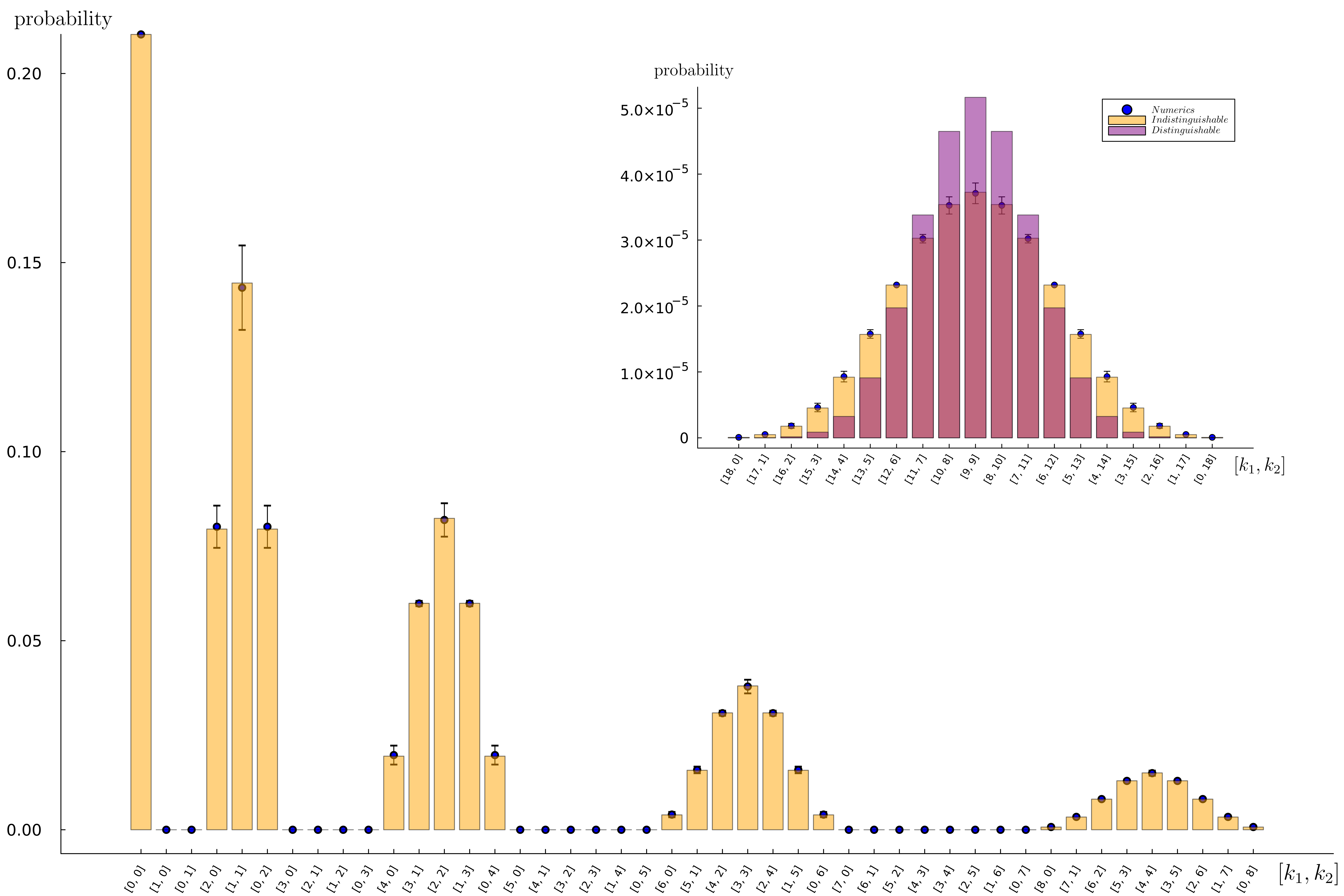}
    \caption{\textbf{Haar-averaged binned probability distribution and asymptotic law}.
    We consider an ideal $m=20$ mode GBS instance where the input states are identical squeezed vacuum states with squeezing parameter $r=0.4$ and the PNR detectors are evenly split into $B=2$ bins. 
    In this figure, we plot the Haar-averaged binned probability distribution of such GBS experiment.
    On the x-axis are the photon count patterns $[k_1,k_2]$ , while on the y-axis are their probabilities.
    The blue dots represent a numerical Haar average over $100$ iterations, their respective error bar representing the standard deviation of the numerical average, while the orange columns follow the asymptotic law given in Eq. \eqref{eq:asymptotics_gaussian_full}. 
    The main figure shows data points up to $8$ detected photons, while the inset shows the data post-selected on measuring $18$ photons.  
    In the insert, we also show the asymptotic law for distinguishable particles in purple.
    Note how only even numbers of photons are detected, consequence of the well known fact that the expansion of the squeezed vacuum state on the Fock basis does not contain odd contributions.}
    \label{fig:asymptotics}
\end{figure*}

Coming back to GBS, let us consider, as an example, the paradigmatic instance of the task where $m$ identical squeezed vacuum input states $\hat{S}(r)\ket{0}$ enter the LON.
Upon post-selecting on detection events with $n$ total photons, the probability of observing a specific pattern $\bm{k}$, averaged over Haar-random unitaries representing the ideal LON, will be approximated by Eq.~\eqref{gaussian_law}, suitably renormalized according to the total photon number distribution.
The latter is equal to $P_m(n/2)$, i.e. the probability of detecting $n/2$ photon pairs, given by  
Eq.~\eqref{negative_binomial}. 
Putting everything together we obtain the Haar-averaged asymptotic law for binned-detector GBS
\begin{align}
    \label{eq:asymptotics_gaussian_full}
    &\expval{P^{\sigma}_{\text{Gaussian}}(\bm{k})} = 
    P_m \left({n}/{2}\right) \expval{P^{\sigma}_{\text{Fock}}(\bm{k})} \\ 
    &= \binom{\frac{m}{2}+\frac{n}{2}-1}{\frac{n}{2}}(\sech{r})^m(\tanh{r})^{n} \expval{P^{\sigma}_{\text{Fock}}(\bm{k})}  \nonumber  \, .
\end{align}
Notice how the above equation holds for even $n$, while $\expval{P^{\sigma}_{\text{Gaussian}}(\bm{k})} = 0$ otherwise, due to the fact that ideal squeezed vacuum states only contain an even number of photons.
We display this distribution in Fig. \ref{fig:asymptotics}, with a comparison to numerical averages. 

\section{Classical mock-up distributions}
\label{sec:mock_up}
Within the context of validating a Gaussian boson sampler, it is of great importance to ensure that the experimental samples are statistically more compatible with the theoretical ground truth of a (possibly lossy) GBS instance, rather than with the output probability distribution of a sampling task that may be efficiently simulated on a classical machine.
This situation arises, for example, when the input states entering the LON are $P$-classical states, i.e. their Glauber-Sudarshan $P$ representation is non-negative.
These classical input states can then be chosen to resemble a squeezed vacuum state.

We remind the reader that any $m$-mode quantum state $\hat{\rho}$ admits a diagonal representation on the coherent state basis by means of the Glauber-Sudarshan $P$ function, namely
\begin{equation}
    \hat{\rho} = \int d^{2m}\bm{\beta} P(\bm{\beta}) \ketbra{\bm{\beta}}  \, .
\end{equation}
Despite being normalized, $P(\bm{\beta})$ may diverge more severely than a delta function and, in general, is not positive semi-definite. 
A state $\hat{\rho}$ is said to be $P$-classical, if its $P$ function is positive and well-behaved.

The computation of the characteristic function for a GBS instance employing $P$-classical input states (details can be found in Appendix \ref{appendix_char_function_P_classical}) proceeds similarly to what we have outlined in Section \ref{sec_characteristic_GBS}, the main difference being that we can now exploit the Glauber-Sudarshan $P$ representation, thus eliminating the necessity of resorting to the positive $P$ representation.
This, in turn, results in  the dimension of the phase space being halved. 
Once again, we start from the definition of the characteristic function 
\begin{equation}
    X(\bm{\eta}) = \Tr{\mathcal{E}(\hat{\rho}_{in}) e^{i\bm{\eta}\cdot\hat{\bm{N}}}} \, ,
\end{equation}
where now the quantum state at the output of the LON can be expressed as 
\begin{equation}
    \mathcal{E}(\hat{\rho}_{in}) =\int d^{2m} \bm{\beta} \, P_{in}(\bm{\beta}) \ketbra{L \bm{\beta}} \, ,
\end{equation}
where $P_{in}$ is the $P$ function of the $m$-mode input state. 
Starting from Eq.~\eqref{phase_shift1} and after some algebra, we obtain 
\begin{equation}
    X(\bm{\eta})  = \int d^{2m} \bm{\beta} \, P_{in}(\bm{\beta}) \, e^{\bm{\beta}^\dagger \mathcal{U}^\intercal \bm{\beta}}  \, ,
    \label{characteristic_P_classical}
\end{equation}
where $\mathcal{U} = L^\intercal H L^*$ and $H$ is defined in Eq.~\eqref{H_matrix_def}. 
In what follows, we focus on two relevant classes of classical input states, namely thermal states and squashed states.
These constitute two of the most common choices of $P$-classical input states that are tested against experimental data coming from a Gaussian boson sampler \cite{martinez2023classical}.

Let us consider an $m$-mode thermal input state $\hat{\rho}_{in} = \otimes_{j=1}^m \hat{\nu}_{th}(\overline{n}_i)$, where $\overline{n}_i$ is the mean photon number of $\hat{\nu}_{th}(\overline{n}_i)$.
One can show that its $P$ function reads 
\begin{equation}
    P_{th}(\bm{\beta}) = \mathcal{N} e^{-\bm{\beta}^\dagger D \bm{\beta}}  \, ,
    \label{thermal_P_function}
\end{equation}
where $D=\text{diag}((\overline{n}_1)^{-1},\dots,(\overline{n}_m)^{-1})$ and
\begin{equation}
    \mathcal{N} = \prod_{i=1}^m \left[\frac{1}{\pi\overline{n}_i}\right] \, .
\end{equation}
We can now substitute Eq.~\eqref{thermal_P_function} into Eq.~\eqref{characteristic_P_classical}, carry out a multi-dimensional Gaussian integral, and obtain 
\begin{equation}
    X(\bm{\eta}) =  \mathcal{N}\frac{(2\pi)^m}{\sqrt{\det{Q}}}  \, ,
\end{equation}
where $Q$ is a complex symmetric matrix defined as 
\begin{equation}
    Q = 
    \begin{pmatrix}
        2D-\mathcal{U}-\mathcal{U}^\intercal & i(\mathcal{U}-\mathcal{U}^\intercal) \\ i(\mathcal{U}^\intercal-\mathcal{U}) & 2D-\mathcal{U}-\mathcal{U}^\intercal
    \end{pmatrix} \, .
\end{equation}

Let us now focus on squashed states, i.e. Gaussian states that exhibit vacuum fluctuations in one quadrature (and more than vacuum fluctuation in its conjugate).
They may conveniently be parametrized as squeezed thermal states, namely
\begin{equation}
    \hat{\rho}_{in} = \bigotimes_{i=1}^m \hat{S}(r_i)\hat{\nu}_{th}({\overline{n}_i})\hat{S}^\dagger(r_i) \, ,
    \label{squashed_state}
\end{equation}
with $\overline{n}_i = (e^{2r_i}-1)/2$.
The $P$-function of the state Eq.~\eqref{squashed_state} reads
\begin{equation}
    P_{sq}(\bm{\beta}) = \mathcal{N} e^{-\bm{x}^\intercal D \bm{x}}\delta^{(m)}(\bm{y}) \, ,
    \label{squashed_P_function}
\end{equation}
where $\bm{\beta} = \bm{x} + i\bm{y}$, $D=\text{diag}(\frac{2}{\lambda_1},\dots,\frac{2}{\lambda_m})$, $\lambda_i = e^{4r_i}-1$ and 
\begin{equation}
    \mathcal{N} = \prod_{i=1}^m \left[\sqrt{\frac{2}{\pi \lambda_i}}\right] \, .
\end{equation}
Substituting Eq.~\eqref{squashed_P_function} into Eq.~\eqref{characteristic_P_classical} and integrating over the $m$-dimensional delta function leads to 
\begin{equation}
    X(\bm{\eta})   = \mathcal{N} \int d^{m}\bm{x}\, e^{-\bm{x}^\intercal (D-\mathcal{U}^\intercal) \bm{x}} \, . 
\end{equation}
Finally, standard multi-dimensional Gaussian integration yields 
\begin{equation}
    X(\bm{\eta}) = \mathcal{N} \sqrt{\frac{(2\pi)^m}{\det{Q}}} \, ,
\end{equation}
with $Q = 2D-\mathcal{U}-\mathcal{U}^\intercal$.

\section{Partial distinguishability}
\label{sec:partial_distinguishability}

\begin{figure}[t]
  \centering
    \includegraphics[width=0.5\textwidth]{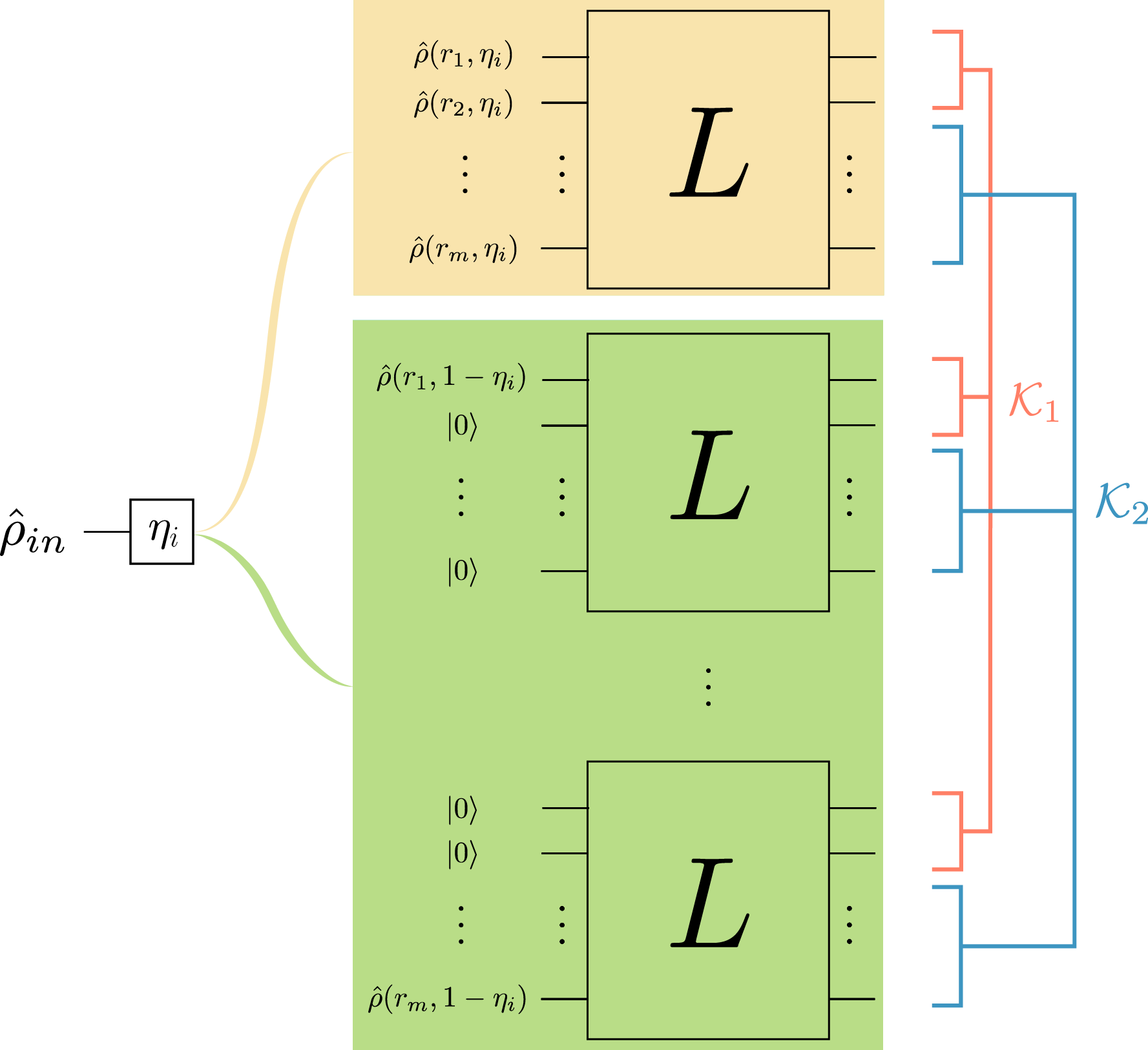}
    \caption{\textbf{
Partial distinguishability model}. Before entering the LON, the input state $\hat{\rho}_{in} = \bigotimes_{i=1}^m\hat{S}(r_i)\ket{0}$ undergoes a process that turns some of the photons into distinguishable ones, and whose probability is parametrized by the indistinguishability efficiency $\eta_i$. 
As a result, the state is split into a indistinguishable component (yellow), and a perfectly distinguishable one (green) that propagates through the LON via $m$ virtual modes without interfering with other photons.
Since the modes contribute independently to the final photo-count, the model is equivalent to simulating $m+1$ independent lossy GBS instances and grouping the corresponding detectors across all modes to retrieve the binned photon-counts (in the above example, we have considered a bi-partition $\lbrace \mathcal{K}_1,\mathcal{K}_2\rbrace$ of the detectors).
Here, $\hat{\rho}(r,\eta)$ denotes a lossy squeezed vacuum state, whose covariance matrix is given by Eq.~\eqref{lossy_squeezed_covariance}.}
    \label{fig:partial}
\end{figure}

\begin{figure*}
    \centering
    \includegraphics[width = 0.9\textwidth]{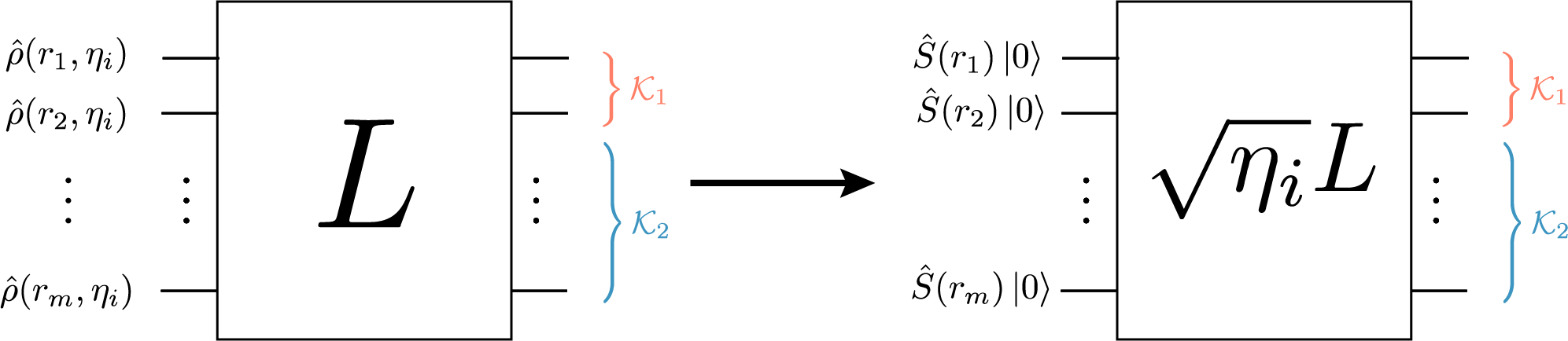}
    \caption{\textbf{Loss absorption into the LON.} On the left, a GBS instance where the LON is described by the subunitary matrix $L$ and is fed with lossy squeezed vacuum states $\hat{\rho}(r,\eta_i)$, whose covariance matrix $\tilde{\sigma}(r,\eta_i)$ is given by Eq.~\eqref{lossy_squeezed_covariance}.
    On the right, an equivalent instance of the task where the initial states' losses have been absorbed into the LON. As a result, the latter is described by the matrix $\sqrt{\eta_i}L$ and is fed with ideal squeezed vacuum states.
    }
    \label{fig:loss_equivalency}
\end{figure*}

Together with losses and detection inefficiencies, partial distinguishability constitutes another source of imperfection that might prevent the sampling task from entering a regime where quantum advantage is, in principle, attainable.
In Ref.~\cite{shi2022effect}, the authors introduced a simple toy model of GBS that aims at capturing some of the phenomenology associated with partial distinguishability of the photons, and studied how the latter $-$ measured by the indistinguishability efficiency $0\leq\eta_i\leq1$ $-$ affects the computational complexity of the problem.
The idea that underlies the model is that, before entering the interferometer, the initially indistinguishable light undergoes a process which turns some of the photons into distinguishable ones.
These then propagate through the LON via virtual modes without interfering with other photons, before being eventually measured by the detectors. 
In the following, we adopt the nomenclature of Ref.~\cite{shi2022effect} and call ``port'' what we have referred to as ``mode'' up until now. 
In fact, each of the LON's $m$ ports is simultaneously populated by the indistinguishable mode as well as by other $m$ additional distinguishable modes, that independently contribute to the photo-count.
Equivalently, each mode spans $m$ ports and we reserve the superscript $(j)$ for quantities related to the $j$-th distinguishable mode.
The indistinguishable mode is initially populated with $m$ squeezed vacuum states.
On the other hand, each virtual mode is initialized in the vacuum state, until a fictitious beam-splitter-like transformation causes the exchange of photons between the $j$-th port of the indistinguishable mode and the $j$-th port of the $j$-th distinguishable mode.

As a result, before entering the LON, both the indistinguishable and distinguishable modes are populated by lossy squeezed vacuum states, their covariance matrices respectively given by (see Ref.~\cite{shi2022effect} for more details)
\begin{equation}
    \sigma = \bigoplus_{j=1}^m \tilde{\sigma}(r_j,\eta_i) \, ,
    \label{cov_indist}
\end{equation}
\begin{equation}
    \sigma^{(j)} = \mathbb{I}_{2j-2} \oplus 
    \tilde{\sigma}(r_j,1-\eta_i)
    \oplus \mathbb{I}_{2m-2j} \, ,
     \label{cov_dist}
\end{equation}
where
\begin{equation}
    \tilde{\sigma}(r,\eta) = 
    \begin{pmatrix}
        \eta e^{2r} +1-\eta & 0 \\ 0 & \eta e^{-2r} +1-\eta 
    \end{pmatrix}
    \label{lossy_squeezed_covariance}
\end{equation}
is the covariance matrix of $\hat{\rho}(r,\eta)$, namely a single-mode squeezed vacuum state $\hat{S}(r)\ket{0}$ that propagated through a loss channel with trasmissivity $\eta$.
The output detection pattern is obtained simply by summing the contribution of the indistinguishable mode $\bm{n}=(n_1,\dots,n_m)$ and those of the virtual modes  $\bm{n}^{(j)}=(n^{(j)}_1,\dots,n^{(j)}_m)$.
As the modes contribute independently to the photo-count, it follows that the model we have described is equivalent to simulating $m+1$ distinct lossy GBS instances where the Gaussian input states have covariance matrices Eq.~\eqref{cov_indist} and Eq.~\eqref{cov_dist}, each $m-$port LON is described by the matrix $L$, and corresponding output ports across the $m+1$ modes are grouped together.
Later we will absorb the fictitious losses of the input state into the LON, in order to retain a GBS implementation whose input ports are fed either with squeezed vacuum or vacuum states.
Detector binning is easily incorporated into this framework by considering a partition of the output ports $\lbrace\mathcal{K}_j \rbrace_{j=1}^B$ across all modes and further grouping together corresponding bins, as can be seen in Figure \ref{fig:partial} (notice how the total number of bins remains $B$).
This is clearly equivalent to simulating a bigger GBS instance with $m(m+1)$ ports however, since most of them are fed with vacuum states, this is not reflected in an increase in the computational complexity of computing the binned probabilities.
We illustrate this with a simple example of an $m-$port GBS experiment with input state $\hat{\rho}_{in} = \ketbra{\psi_{in}}$, where $\ket{\psi_{in}} = \hat{S}(r)\!\ket{0}\otimes \ket{0}^{\otimes m-1}$.
Exploiting the positive representation on the phase real space of a squeezed vacuum state we can write 
\begin{equation}
    \hat{\rho}_{in} = \int d^m\bm{x} \, d^m\bm{y} \, P_{in}(\bm{x},\bm{y})\frac{\ket{\bm{x}}\!\!\bra{\bm{y}}}{\braket{\bm{y}\vert\bm{{x}}}} \, , 
\end{equation}
where $ P_{in}(\bm{x},\bm{y}) = P({x}_1,{y}_1) \prod_{i=2}^m \delta(x_i) \delta(y_i)$ and $P(x,y)$ is defined in Eq.~\eqref{positive_P_squeezed_vacuum}.
The characteristic function is given by
\begin{equation}
    X(\bm{\eta}) = \int d^m\bm{x} \, d^m\bm{y} \, P_{in}(\bm{x},\bm{y}) e^{\bm{x}^\intercal \mathcal{U} \bm{y}} \, ,
\end{equation}
and after integrating over the delta functions we are left with the two-dimensional Gaussian integral
\begin{equation}
    X(\bm{\eta}) = \frac{\sqrt{1+\gamma}}{\pi\gamma} \int d^2\bm{z}  e^{-\frac{1}{2}\bm{z}^\intercal Q \bm{z}} = \frac{2\sqrt{1+\gamma}}{\gamma\sqrt{\det{Q}}} \, ,
\end{equation}
where $\bm{z}=(x_1,y_1)$ and 
\begin{equation}
    Q = 
    \begin{pmatrix}
        2\gamma^{-1}+1 & -1-\mathcal{U}_{11} \\ 
        -1-\mathcal{U}_{11} & 2\gamma^{-1}+1
    \end{pmatrix} \, .
\end{equation}
Hence, in this scenario, computing the characteristic function amounts to the calculation of the determinant of a $2\times 2$ matrix, while in Section \ref{sec_characteristic_GBS} we showed that when all input ports are fed with squeezed light the $Q$ matrix is $2m\times 2m$.
One then easily realizes that, in general, each vacuum input state reduces the dimension of the $Q$ matrix by 2.

Consequently, computing the characteristic function of a GBS instance with partial distinguishability amounts to computing the determinant of a $4m\times 4m$ matrix, thus retaining the same scaling of the ideal indistinguishable case. 
Lastly, we point out that the fictitious losses of the input states may be absorbed into the state's evolution, as depicted in Figure \ref{fig:loss_equivalency}.
In particular, we consider a bigger LON that describes the evolution of all ports across all modes, characterized by the matrix 
\begin{equation}
    \tilde{L} = \sqrt{\eta_i} L 
    \oplus (\sqrt{1-\eta_i}L)^{\oplus m} \, .
\end{equation}
Note that simulating GBS with fully distinguishable photons may be achieved by setting $\eta_i = 0$, which causes the indistinguishable mode to disappear. In this case, the big LON is described by 
\begin{equation}
    \tilde{L} = \bigoplus_{i=1}^m L \, .
\end{equation}
Hence, following the argument above, computing the characteristic function in this scenario amounts to the calculation of the determinant of a $2m\times 2m$ matrix, same as for the ideal indistinguishable case.

\section{Conclusions}
\label{sec:conclusions}
In this paper we have studied the problem of validating a Gaussian boson sampler via detector binning.
In particular, we showed how to compute the binned-count probability distribution for a GBS instance employing PNR detectors, by means of discrete Fourier transform of the related characteristic function.
We derived an analytical closed formula for the latter and showed that its computation only involves the evaluation of matrix determinants, thus ensuring the computational efficiency of the protocol.
Our method can accommodate for multiple noise sources, including loss and partial distinguishability.
This is a crucial requirement to substantiate any claims of quantum advantage, as the presence of noise and imperfections may render the task classically efficiently simulable, thus preventing it from reaching the regime where quantum speedup is achievable.
Additionally, our method encompasses known validation techniques based on marginal probabilities and correlation functions, and may also be used to compute binned-count probability distributions for classical inputs such as thermal states and squashed states. 
These can then be used to certify that experimental samples are statistically more compatible with the ground truth of GBS, rather then with an efficiently computable probability distribution that only resembles the latter.
Lastly, we computed Haar averages of the binned-count probability distribution for an ideal GBS task and showed that, at fixed number of detected photons, one obtains a Gaussian profile in the asymptotic limit. 
Given the versatility of the method, we believe that detector binning is a practical and computationally inexpensive way to help substantiate claims of quantum computational advantage in future GBS experiments.

\section{Acknowledgments}
G.B. is part of the AppQInfo MSCA ITN which received funding from the European Union’s Horizon 2020 research and innovation programme under the Marie Sklodowska-Curie grant agreement No 956071. 
B.S. is a Research Fellow of the Fonds National de la Recherche Scientifique – FNRS. 
MSK acknowledges the Samsung GRC programme and the UK EPSRC through EP/W032643/1 and EP/Y004752/1. L.N. acknowledges support from FCT-Fundação para a Ciência e a Tecnologia (Portugal) via the Project No. CEECINST/00062/2018 and from the European Union's Horizon 2020 research and innovation program through the FET project PHOQUSING (“PHOtonic QUantum SamplING machine” - Grant Agreement No. 899544).
B.S. thanks Ursula Cardenas Mamani for help on the figures. 

\bibliographystyle{unsrtnat}
\bibliography{biblio}

\onecolumn
\newpage
\appendix

\section{Binned probability and characteristic function}
\label{appendix_characteristic_expectation}
Here, we explicitly show that the two expressions of the characteristic function $X(\bm{\eta})$ introduced in the main text Eq.~\eqref{characteristic_function_definition_1} and Eq.~\eqref{characteristic_function_definition_2}, indeed coincide. 
Expanding the state $\hat{\rho}$ on the Fock state basis 
\begin{equation}
    \ket{\bm{s}} \equiv \bigotimes_{j=1}^m \frac{\hat{a}^{\dagger s_j}_j}{\sqrt{s_j!}}\ket{0} \, 
\end{equation}
yields
\begin{equation}
\begin{split}
     X(\bm{\eta}) & = \Tr\lbrace \hat{\rho} \, e^{i\bm{\eta}\cdot\hat{\bm{N}}}\rbrace = \sum_{\bm{s},\bm{t}}\Tr\lbrace\ketbra{\bm{s}}\hat{\rho}\ketbra{\bm{t}}e^{i\bm{\eta}\cdot\hat{\bm{N}}}\rbrace \\ &
     = \sum_{\bm{s},\bm{t}}\bra{\bm{s}}\hat{\rho}\ketbra{\bm{t}}e^{i\bm{\eta}\cdot\hat{\bm{N}}}\ket{\bm{s}} 
     \\  & = 
     \sum_{\bm{s},\bm{t}}\bra{\bm{s}}\hat{\rho}\ketbra{\bm{t}}e^{i\sum_j \eta_j \sum_{\ell\in\mathcal{K}_j}\hat{n}_\ell}\ket{\bm{s}}  
     \\ & 
      = \sum_{\bm{s}}\bra{\bm{s}}\hat{\rho}\ket{\bm{s}}e^{i\sum_j \eta_j \sum_{\ell\in\mathcal{K}_j}s_\ell} \, .
\end{split}
\end{equation}
Crucially, the sum over $\bm{s}$ can now be decomposed as a sum over all possible partitioned modes' detection patterns $\bm{k}$, and a sum over the Fock states $\ket{\bm{s}}$ that are compatible with $\bm{k}$, i.e. $k_j = \sum_{\ell\in\mathcal{K}_j}s_\ell$ for every $j\in\lbrace 1,\dots,B\rbrace$.
Hence, we obtain
\begin{equation}
    \begin{split}
     X(\bm{\eta}) & = 
     \sum_{\bm{k}}\sum_{\bm{s}\vert\bm{k}}\bra{\bm{s}}\hat{\rho}\ket{\bm{s}} e^{i\sum_j \eta_j \sum_{\ell\in\mathcal{K}_j}s_\ell}
     \\ & = \sum_{\bm{k}} e^{i\bm{\eta}\cdot\bm{k}} \sum_{\bm{s}\vert\bm{k}}\bra{\bm{s}}\hat{\rho}\ket{\bm{s}} = 
     \sum_{\bm{k}} P(\bm{k}) e^{i\bm{\eta}\cdot\bm{k}} \, ,
\end{split}
\end{equation}
where we have used the fact that $\sum_{\ell\in\mathcal{K}_j}s_\ell = k_j$ and that $\sum_{\bm{s}\vert\bm{k}}\bra{\bm{s}}\rho\ket{\bm{s}}= P(\bm{k})$.

\section{Characteristic function for GBS}
\label{appendix_detailed_calculation}
A GBS experiment consists of injecting a LON with squeezed vacuum states and sampling from the output photon-number distribution.
The $m-$mode initial state $\hat{\rho}_{in}$ thus reads
\begin{equation}
    \hat{\rho}_{in} =  \bigotimes_{j=1}^m \hat{S}(r_j)\ketbra{0}\hat{S}^\dagger (r_j) \, ,
\end{equation}
where 
\begin{equation}
    \hat{S}(r_j) = e^{{\frac{r_j}{2}(\hat{a}_j^{\dagger 2}-\hat{a}_j^2)}}
\end{equation}
is the single-mode squeezing operator and $r_j>0$ is the squeezing parameter.
The quantum evolution of a state via the lossy LON is described by the CP-map $\mathcal{E}$. 
We recall that the corresponding transformation of the system's modes is linear, hence it is fully characterized by a sub-unitary matrix $L$, i.e. $L^\dagger L \leq \mathbb{I}_m$, with $\mathbb{I}_m$ denoting the $m\times m$ identity matrix. 
Hence, given a partition $\lbrace\mathcal{K}_j\rbrace_{j=1}^B$  of the output modes into $B$ bins, the characteristic function of a GBS experiment we aim to compute reads
\begin{equation}
    X(\bm{\eta}) = \Tr{\mathcal{E}(\hat{\rho}_{in}) e^{i\bm{\eta}\cdot\hat{\bm{N}}}} \, , 
    \label{characteristic_function_definition_GBS}
\end{equation}
where
\begin{equation}
    \bm{\eta}\cdot\hat{\bm{N}} = \sum_{j=1}^B \eta_ j\sum_{\ell\in\mathcal{K}_j}\hat{n}_\ell \, .
\end{equation}
In Ref.~\cite{oh2022quantum} the authors computed this quantity for an ideal system, where the evolution is described by a unitary matrix. Here, we generalize the calculation by first allowing noisy evolution described by $L$ and in Appendix \ref{appendix_char_function_P_classical} we compute the characteristic function for $P-$classical states.

Using the identity 
\begin{equation}
   e^{i\theta\hat{n}} = :e^{(e^{i\theta}-1)\hat{n}}: 
   \label{identity_normal_ordering}
\end{equation}
we can write the phase-shift operator appearing in Eq.~\eqref{characteristic_function_definition_GBS} as
\begin{equation}
    e^{i\bm{\eta}\cdot\hat{\bm{N}}} = \bigotimes_{j=1}^B \bigotimes_{\ell\in\mathcal{K}_j}e^{i\eta_j \hat{n}_\ell} = \bigotimes_{j=1}^B \bigotimes_{\ell\in\mathcal{K}_j} :e^{(e^{i\eta_j}-1)\hat{n}_\ell} : \, .
    \label{phase_shift}
\end{equation}
We can prove Eq.~\eqref{identity_normal_ordering} by explicit computation of the matrix elements of the two operators on the coherent state basis $\ket{\alpha}$. In particular, we obtain
\begin{equation}
\begin{split}
     \bra{\alpha}e^{i\theta\hat{n}}\ket{\alpha}   = \sum_{n=0}^{\infty} \bra{\alpha}e^{i\theta \hat{n}} \ket{n}\!\!\braket{n\vert\alpha} = \sum_{n=0}^{\infty} e^{i\theta n}\, \vert\!\braket{n\vert\alpha}\!\vert^2 =
   e^{-\vert\alpha\vert^2} \sum_{n=0}^{\infty}   \frac{e^{i\theta n}\vert\alpha\vert^{2n}}{n!} = e^{(e^{i\theta}-1)\vert\alpha\vert^2} \, ,
\end{split}
\label{identity_proof_2}
\end{equation}
and
\begin{equation}
\begin{split}
  \bra{\alpha}:e^{(e^{i\theta}-1)\hat{n}}:\ket{\alpha}  =\bra{\alpha}:e^{(e^{i\theta}-1)\hat{a}^\dagger\hat{a}}:\ket{\alpha} = e^{(e^{i\theta}-1)\vert\alpha\vert^2} 
\end{split}
\label{identity_proof_1}
\end{equation}
where we have used the representation of the coherent state on the Fock state basis, namely 
\begin{equation}
    \ket{\alpha}=e^{-\frac{\vert\alpha\vert^2}{2}}\sum_n \frac{\alpha^n}{\sqrt{n!}}\ket{n} \, .
\end{equation}
Eq.~\eqref{identity_proof_1} and Eq.~\eqref{identity_proof_2} manifestly coincide, thus concluding our proof.
\\
\\
The expectation value of a normally-ordered operator may then be evaluated by averaging over the phase-space variables according to any generalized $P$ distribution of the state.
A generic $m-$mode quantum state $\hat{\rho}$ admits a non-singular phase-space representation via a quasi-probability distribution $P(\bm{\alpha},\bm{\beta})$ such that
\begin{equation}
    \hat{\rho} = \int_{{\mathbb{C}^{2m}}} P(\bm{\alpha,\bm{\beta}})\, \frac{\ket{\bm{\alpha}}\!\!\bra{\bm{\beta}^*}}{\braket{\bm{\beta}^*\vert{\bm{\alpha}}}} \,d\mu(\bm{\alpha,\bm{\beta}}) \, ,
    \label{generalized_P_representation}
\end{equation}
where $\bm{\alpha},\bm{\beta}\in\mathbb{C}^m$ and $\ket{\bm{\alpha}} = \ket{\alpha_1}\otimes \cdots \otimes \ket{\alpha_m}$ is an $m-$mode coherent state. 
The specific functional form of $P(\bm{\alpha,\bm{\beta}})$ depends on the choice of the integration measure, and it can be showed that $d\mu (\bm{\alpha},\bm{\beta})=d^{2m}\bm{\alpha}\,d^{2m}\bm{\beta}$ leads to a non-negative representation for any state, which we call positive $P$ representation.
Hence, the output state of the LON $\mathcal{E}(\hat{\rho}_{in})$ can be expressed as 
\begin{equation}
     \mathcal{E}(\hat{\rho}_{in}) = \int_{{\mathbb{C}^{2m}}} P_{in}(\bm{\alpha,\bm{\beta}})\, \frac{\mathcal{E}(\ket{\bm{\alpha}}\!\!\bra{\bm{\beta}^*})}{\braket{\bm{\beta}^*\vert{\bm{\alpha}}}} \,d^{2m}\bm{\alpha}\,d^{2m}\bm{\beta} \, ,
     \label{output_state_P}
\end{equation}
where $P_{in}$ is the positive $P$ representation of $\hat{\rho}_{in}$ and the integral spans the whole $2m-$dimensional complex space, thus corresponding to a $4m-$dimensional real volume integral.
\\
In order to compute the action of the CP-map $\mathcal{E}$ on the operator $\ket{\bm{\alpha}}\!\!\bra{\bm{\beta}^*}$, we recall that an $m-$mode lossy LON may be simply modelled by considering a bigger $2m$-mode loss-less interferometer where the additional $m$ environmental modes are initialized in the vacuum state, and a final trace is taken over the environmental degrees of freedom.
This ideal LON is characterized by a $2m\times2m$ unitary block matrix $T$ that reads
\begin{equation}
    T=\begin{pmatrix}
    L & N \\ P & M 
    \end{pmatrix} \, ,
\end{equation}
and whose unitarity enforces the constraint 
\begin{equation}
    L^\dagger L + P^\dagger P = \mathbb{I}_m \, .
    \label{unitarity}
\end{equation}
We can now compute 
\begin{equation}
\begin{split}
        \mathcal{E}{(\ket{\bm{\alpha}}\!\!\bra{\bm{\beta}^*})}  & = \Tr_{env}\lbrace\hat{T} \left(\ket{\bm{\alpha}}\!\!\bra{\bm{\beta}^*}\otimes\ketbra{\bm{0}} \right) \hat{T}^\dagger\rbrace
      = \Tr_{env}\lbrace\ket{T(\bm{\alpha}\,\bm{0})}\!\!\bra{T(\bm{\beta}^*\,\bm{0})}\rbrace
    \\ & = \Tr_{env}\lbrace\ket{L\bm{\alpha}\,P\bm{\alpha}}\!\!\bra{L\bm{\beta}^*\,P\bm{\beta}^*}\rbrace  = \braket{P\bm{\beta}^*\vert P\bm{\alpha}} \ket{L\bm{\alpha}}\!\!\bra{L\bm{\beta}^*} \, , 
    \label{action_on_coherent}
\end{split}
\end{equation}
where $\hat{T}$ is the unitary operator that describes the loss-less LON and the trace is taken over the $m$ environmental modes.
Substituting Eq.~\eqref{phase_shift}, Eq.~\eqref{output_state_P} and Eq.~\eqref{action_on_coherent} into Eq.~\eqref{characteristic_function_definition_GBS} yields
\begin{equation}
\begin{split}
     X(\bm{\eta}) &  =
     \int_{{\mathbb{C}^{2m}}}{P_{in}(\bm{\alpha},\bm{\beta})}\frac{\braket{P\bm{\beta}^*\vert P\bm{\alpha}}}{\braket{\bm{\beta}^*\vert\bm{\alpha}}}
    \bra{L\bm{\beta}^*} \bigotimes_{j=1}^B \bigotimes_{\ell\in\mathcal{K}_j} :e^{(e^{i\eta_j}-1)\hat{n}_\ell} : \ket{L\bm{\alpha}} \,d^{2m}\bm{\alpha}\,d^{2m}\bm{\beta}
    \\  &  =\int_{{\mathbb{C}^{2m}}}P_{in}(\bm{\alpha},\bm{\beta})
 \frac{\braket{P\bm{\beta}^*\vert P\bm{\alpha}}\braket{L\bm{\beta}^*\vert L\bm{\alpha}}}{\braket{\bm{\beta}^*\vert \bm{\alpha}}}
    e^{\sum_{j=1}^B (e^{i\eta_j}-1) \sum_{\ell\in\mathcal{K}_j}(L^*\bm{\beta})_\ell(L\bm{\alpha})_\ell}\,d^{2m}\bm{\alpha}\,d^{2m}\bm{\beta} \, ,
    \label{characteritic_function_integral}
\end{split}
\end{equation}
where we have used the fact that matrix elements on the coherent state basis of normally-ordered operators satisfy the property $\bra{\beta} : f(\hat{a}^\dagger,\hat{a}): \ket{\alpha} = \braket{\beta\vert\alpha} f(\beta^*,\alpha) \, ,$
where $f(\hat{a}^\dagger,\hat{a})$ is a generic function of the bosonic operators.
We can also conveniently rewrite the exponential appearing in the expression above as
\begin{equation} 
     e^{\sum_{j=1}^B (e^{i\eta_j}-1)\sum_{\ell\in\mathcal{K}_j}\left(L^*\bm{\beta}\right)_\ell\left(L\bm{\alpha}\right)_\ell } = e^{\bm{\alpha}^\intercal\mathcal{U}\bm{\beta}} \, ,
     \label{rewritten_exponent}
\end{equation}
where $\mathcal{U} = L^\intercal H L^*$ and $H$ is a diagonal matrix that contains the phase information, defined as 
\begin{equation}
    H=\text{diag}(e^{i\theta_1}-1,\dots,e^{i\theta_M}-1) \, , \quad\text{with}\quad \theta_i = \eta_j\quad\text{if}\quad i\in\mathcal{K}_j \, .
    \label{H_matrix_def}
\end{equation}
Furthermore, using the well known formula for the overlap between coherent states
\begin{equation}
    \braket{\beta\vert\alpha} = e^{-\frac{1}{2}(\vert\beta\vert^2 + \vert\alpha\vert^2 - 2\beta^* \alpha)} \, ,
\end{equation}
one easily proves that
\begin{equation}
\frac{\braket{P\bm{\beta}^*\vert P\bm{\alpha}}\braket{L\bm{\beta}^*\vert L\bm{\alpha}}}{\braket{\bm{\beta}^*\vert \bm{\alpha}}} = e^{-\frac{1}{2}[\bm{\beta}^\intercal (L^\dagger L + P^\dagger P -\mathbb{I})\bm{\beta}^*+\bm{\alpha}^{*\intercal}(L^\dagger L + P^\dagger P -\mathbb{I})\bm{\alpha}-2\bm{\beta}^{\intercal}(L^\dagger L + P^\dagger P -\mathbb{I})\bm{\alpha}]} = 1 \, ,
\end{equation}
where we have used the unitarity constraint Eq.~\eqref{unitarity}.
Hence, the characteristic function reads
\begin{equation}
    X(\bm{\eta})  = 
    \int_{{\mathbb{C}^{2m}}} 
    P_{in}(\bm{\alpha},\bm{\beta}) e^{\bm{\alpha}^\intercal\mathcal{U}\bm{\beta}}d^{2m}\bm{\alpha}\,d^{2m}\bm{\beta} \, .
     \label{characteristic_final_integral}
\end{equation}
A single-mode squeezed vacuum state $S(r)\ket{0}$ admits a positive $P$ representation on a two-dimensional \textit{real} space that reads \cite{drummond2022simulating} 
\begin{equation}
    P(x,y) = \frac{\sqrt{1+\gamma}}{\pi\gamma} e^{-(x^2+y^2)(\gamma^{-1}+{1}/{2})+xy} \, ,
    \label{positive_P_squeezed_vacuum}
\end{equation}
where $1+\gamma = e^{2r}$. Note that the above expression  holds for strictly positive values of the squeezing parameter.  
Since $\hat{\rho}_{in}$ is a tensor product, it follows that $P_{in}(\bm{\alpha},\bm{\beta})$ is simply the product of the positive $P$ distributions of the squeezed vacuum states, i.e.
\begin{equation}
    P_{in}(\bm{x},\bm{y}) = \prod_{i=1}^m \left[ \frac{\sqrt{1+\gamma_i}}{\pi\gamma_i} e^{-(x_i^2+y_i^2)(\gamma_i^{-1}+{1}/{2})+x_i y_i} \right] = \mathcal{N}e^{-\bm{x}^\intercal A \bm{x} - \bm{y}^\intercal A \bm{y} + \bm{x}^\intercal \bm{y}} \, ,
    \label{P_function_initial_state}
\end{equation}
where 
\begin{equation}
    \mathcal{N} = \prod_{i=1}^m \left[ \frac{\sqrt{1+\gamma_i}}{\pi\gamma_i}\right] \, , \quad A=\text{diag}\lbrace \gamma_i^{-1}+{1}/{2} \rbrace_{i=1}^m \, .
\end{equation}
We can now substitute Eq.~\eqref{P_function_initial_state} into Eq.~\eqref{characteristic_final_integral} and  obtain
\begin{equation}
    X(\bm{\eta}) = \mathcal{N} \int_{\mathbb{R}^{2m}} d^m\bm{x}\,d^m\bm{y} \, e^{-\bm{x}^\intercal A \bm{x} - \bm{y}^\intercal A \bm{y} - \bm{x}^\intercal {B}\bm{y} - \bm{y}^\intercal {B}^\intercal \bm{x} } =  \mathcal{N} \int_{\mathbb{R}^{2m}} d^{2m}\bm{z}\,e^{-\frac{1}{2}\bm{z}^\intercal Q \bm{z}} \, ,
    \label{final_gaussian_integral}
\end{equation}
where $B=-(\mathbb{I}+\mathcal{U})/2$, $\bm{z}=(\bm{x},\bm{y})$ and the matrix $Q$ defined as
\begin{equation}
    Q = 2\begin{pmatrix}
        A & B \\ B^\intercal  & A
    \end{pmatrix} = 
    \begin{pmatrix}
        2\Gamma^{-1} + \mathbb{I}_m  & -L^\intercal \text{diag}\lbrace e^{i\theta_j} \rbrace_{j=1}^m L^* \\
        -L^\dagger \text{diag}\lbrace e^{i\theta_j} \rbrace_{j=1}^m L &  2\Gamma^{-1} + \mathbb{I}_m 
        \end{pmatrix}
    \, 
\end{equation}
with $\Gamma = \text{diag}\lbrace \gamma_j \rbrace_{j=1}^m$.
Notice how the positive-definiteness of the real part of the complex symmetric matrix $Q$ ensures the convergence of the Gaussian integral in Eq.~\eqref{final_gaussian_integral}.
Straightforward multi-dimensional Gaussian integration yields the final result
\begin{equation}
   X(\bm{\eta}) = \mathcal{N} \frac{(2\pi)^m}{\sqrt{\det{Q}}}  = \prod_{i=1}^m \left[ \frac{2\sqrt{1+\gamma_i}}{\gamma_i}\right]\frac{1}{\sqrt{\det{Q}}} \, .
\end{equation}

\section{Energy cutoff}
\label{appendix:cutoff}
The sample space size of a GBS experiment employing PNR detection is naturally infinite because of the Gaussian nature of the input states. The latter implies that the number of photons reaching the detectors is not fixed, therefore bringing forth the necessity to introduce an energy cutoff $n$ such that the probability of observing more than $n$ photons in any of the bins is negligible.
This, of course, depends on the specific partition of the output modes, however it is clear that this condition is automatically satisfied if we require the probability of having more than $n$ photons entering the passive LON to be exponentially small.
Notice that if the GBS instance at study is operating in the non-collisional regime, i.e. the probability of observing more then one photon in any output mode is highly suppressed (also an assumption of current complexity proof of GBS), then the total number of photons is much smaller than the number of modes $m$, and we can safely set the cutoff to $n = m$. 
In the following, we focus on the particularly relevant case of identical squeezed vacuum states entering the interferometer.
We emphasize that the LON does not contain active optical elements, meaning that the total number of photons may only decrease due to losses within the system. 
\\
\\
The expansion on the Fock basis of a single-mode squeezed vacuum state
\begin{equation}
    S(r)\ket{0} = \frac{1}{\sqrt{\cosh{r}}} \sum_{n=0}^\infty (\tanh{r})^n \frac{\sqrt{(2n)!}}{2^n n!} \ket{2n} 
\end{equation}
reveals that the latter contains even number of photons only, with the probability of observing $k$ couples of photons reading
\begin{equation}
    P_{1}(k) = \frac{(\tanh{r})^{2k}}{\cosh{r}} \cdot \frac{(2k)!}{(2^k k!)^2} \, .
    \label{squeezed_vacuum_probability}
\end{equation}
If we now consider $m$ identical squeezed vacuum states, then the probability $P_m (k)$ of observing a total of $k$ photon pairs is obtained by subsequent convolution of Eq.~\eqref{squeezed_vacuum_probability}.
In particular, for even $m$ one obtains \cite{GBS_original} 
\begin{equation}
    P_m (k) = \binom{\frac{m}{2}+k-1}{k}(\sech{r})^m(\tanh{r})^{2k} \, ,
    \label{negative_binomial}
\end{equation}
i.e. a negative binomial distribution. 
We recall that the average photon number of $S(r)\ket{0}$ is $(\sinh{r})^2$, hence the mean value of the total photon pairs distribution $P_m$ is simply $\frac{m}{2}(\sinh{r})^2$.  
The extension of Eq.~\eqref{negative_binomial} to odd values of $m$ is achieved by employing the Gamma function to generalize the factorial, namely
\begin{equation}
     P_m (k) = \frac{\Gamma(k+{m}/{2})}{\Gamma ({m}/{2})k!}
     (\sech{r})^m(\tanh{r})^{2k}\, .
\end{equation}
The negative binomial distribution with support on the set $\lbrace 0,1,2,\dots\rbrace$ models the number of observed failures witnessed before $n$ successes in consecutive Bernoulli trials. Hence, if $Y_n\sim \text{NB}(n,p)$ then 
\begin{equation}
    \mathbb{P}[Y_n=k] = \binom{k+n-1}{k} (1-p)^k p^n \, ,
    \label{negative_binomial_wiki}
\end{equation}
where $p$ is the success probability of a single Bernoulli trial. 
By comparing Eq.~\eqref{negative_binomial} with the parametrization of Eq.~\eqref{negative_binomial_wiki}, we establish the correspondences 
\begin{equation}
    n=m/2 \, , \quad p=(\sech{r})^2 \, , \quad 1-p=(\tanh{r})^2 \, .
    \label{identifications}
\end{equation}
Let $B_{s+n}$ be a random binomial variable with $s+n$ and $p$ being the number of trials and the success probability, respectively.
The following identity holds
\begin{equation}
    \mathbb{P}[Y_n > s] = \mathbb{P}[B_{s+n}<n] \, ,
    \label{negbin_to_bin}
\end{equation}
i.e. the probability of observing more than $s$ failures before having witnessed $n$ successes is equal to the probability of observing less than $n$ successes in $s+n$ trials.
\\
Our aim is to derive an anti-concentration inequality for the negative binomial distribution, i.e. we want to bound 
\begin{equation}
    \mathbb{P}[Y_n>\alpha \mathbb{E}[Y_n]] = \mathbb{P}[B_{\alpha\mathbb{E}[Y_n]+n}<n] \, ,
    \label{useful_identity}
\end{equation}
where $\alpha>1$.
The equation above reveals that it is possible to bound the tail of the negative binomial distribution by exploiting the properties of the binomial distribution.
The expectation value of $B_{\alpha\mathbb{E}[Y_n]+n}$ reads
\begin{equation}
    \mathbb{E}[B_{\alpha\mathbb{E}[Y_n]+n}] = (\alpha\mathbb{E}[Y_n]+n)p  = 
    n(\alpha (1-p)  + p) \, ,
\end{equation}
hence we can write Eq.~\eqref{useful_identity} as
\begin{equation}
     \mathbb{P}[Y_n>\alpha \mathbb{E}[Y_n]] = 
     \mathbb{P}\left[B_{\alpha\mathbb{E}[Y_n]+n}<\frac{  \mathbb{E}[B_{\alpha\mathbb{E}[Y_n]+n}]}{\alpha(1-p)+p}\right] \, .
\end{equation}
The equation above reveals that it is possible to bound the tail of the negative binomial distribution by exploiting Chernoff's bound for the binomial distribution's lower tail \cite{concentration_inequalities} 
\begin{equation}
    \mathbb{P}[B<(1-\varepsilon)\mathbb{E}[B]]\leq \text{exp}\left({-\frac{\varepsilon^2}{2}\mathbb{E}[B]} \right) \, ,
    \label{chernoff}
\end{equation}
where $B$ is a generic binomial random variable and $0<\varepsilon<1$.
In particular, using the parameter identifications in Eq.~\eqref{identifications} and $1-\varepsilon = (\alpha(1-p)+p)^{-1}$ we obtain the bound we were looking for, namely
\begin{equation}
    \mathbb{P}\left[k> \frac{\alpha m \sinh^2{r}}{2}\right] \leq \text{exp} \left( -m \frac{(\alpha-1)^2\sinh^2{r} \tanh^2{r}}{4(1+\alpha\sinh^2{r})}\right) \, . 
    \label{final_cutoff_bound}
\end{equation}
This bound is exponentially decreasing in the number of modes $m$ and that any accuracy can be achieved by tuning $\alpha$.
In particular, as $\alpha$ increases, the truncation error decreases exponentially.This is expected since, in the limit of many modes $m\gg 1$, the total photon number distribution distribution Eq.~\eqref{negative_binomial} converges to a normal distribution by virtue of the central limit theorem.
Consequently, in this regime, one could safely replace Eq.~\eqref{final_cutoff_bound} with suitable bounds for the tail of a Gaussian.
\\
\\
In order to derive a cutoff photon number ensuring a target error $\epsilon$, let us reparametrize $\alpha$ as
\begin{equation}
    \alpha= 2+ \frac{\lambda}{\sinh^2{r}},
    \label{def_lambda}
\end{equation}
for $\lambda\geq 1$. By assumption, we have that $\alpha \sinh^2{r}\geq 1$. Using this in Eq.~\eqref{final_cutoff_bound}, we can write 
\begin{equation}
    \mathbb{P}\left[k> \frac{\alpha m \sinh^2{r}}{2}\right] \leq \text{exp} \left( -m \frac{(\alpha-1)^2\tanh^2{r}}{8\alpha}\right) \, . 
    \label{cutoff_bound_2}
\end{equation}
Moreover, using the fact that $(\alpha-1)^2\geq \alpha (\alpha-2)$ together with \eqref{def_lambda} we obtain a simpler bound
\begin{equation}
    \mathbb{P}\left[k> m \sinh^2{r} + \frac{ m \lambda}{2}\right] \leq \text{exp} \left( -\frac{\lambda m}{8\cosh^2{r}}\right) \, . 
    \label{cutoff_bound_3}
\end{equation}
This implies that if we choose a cutoff photon number
\begin{equation}
    n= m \sinh^2(r) + 4 \cosh^2{r}\log\left(\frac{1}{\epsilon}\right)
\end{equation}
we can ensure that  $\mathbb{P}\left[k>n\right] \leq \epsilon$.

\section{Characteristic function for $P$-classical input states}
\label{appendix_char_function_P_classical}
In this Appendix we compute the characteristic function of a sampling experiment where thermal or squashed states are sent into a LON described by sub-unitary matrix $L$, before being measured by PNR detectors that are grouped into bins according to the partition $\lbrace \mathcal{K}_j\rbrace_{j=1}^B$.
\\
The calculations closely follow those presented in Appendix \ref{appendix_detailed_calculation}, the main difference being that the classical nature of these two classes of input states at study allows us to exploit their Glauber-Sudarshan $P$ representation rather than their positive $P$ representation, effectively halving the dimension of the phase space.
Any $m$-mode quantum state $\hat{\rho}$ admits a diagonal representation on the coherent state basis
\begin{equation}
    \hat{\rho} = \int d^{2m} \bm{\beta} P(\bm{\beta}) \ketbra{\bm{\beta}} \, ,
\end{equation}
where $P(\bm{\beta})$ is the state's $P$ function.
The latter typically displays negativities and severe divergencies, however it is well defined and positive-definite for states $-$ like thermal and squashed $-$ that lack genuine quantum properties. 
Following Appendix \ref{appendix_detailed_calculation}, we start from 
\begin{equation}
    X(\bm{\eta}) = \Tr{\mathcal{E}(\hat{\rho}_{in}) e^{i\bm{\eta}\cdot\hat{\bm{N}}}}  = \Tr{\mathcal{E}(\hat{\rho}_{in})  \bigotimes_{j=1}^B \bigotimes_{\ell\in\mathcal{K}_j} :e^{(e^{i\eta_j}-1)\hat{n}_\ell} : }\, , 
\end{equation}
and exploit the $P$-function representation of the initial state $\hat{\rho}_{in}$ to express the output state as
\begin{equation}
    \mathcal{E}(\hat{\rho}_{in}) = \int d^{2m} \bm{\beta} \, P_{in}(\bm{\beta})  \mathcal{E} (\ketbra{\bm{\beta}})  = \int d^{2m} \bm{\beta} \, P_{in}(\bm{\beta}) \ketbra{L \bm{\beta}} \, .
\end{equation}
Here, we have used the fact that, by definition, a LON described by the sub-unitary matrix $L$ sends a multi-mode coherent state $\ket{\bm{\beta}}$ to $\ket{L\bm{\beta}}$. 
Hence, the characteristic function reads
\begin{equation}
\begin{split}
    X(\bm{\eta})  & = \int d^{2m} \bm{\beta} \, P_{in}(\bm{\beta}) \bra{L\bm{\beta}} \bigotimes_{j=1}^B \bigotimes_{\ell\in\mathcal{K}_j} :e^{(e^{i\eta_j}-1)\hat{n}_\ell} : \ket{L\bm{\beta}}
    \\ & = \int d^{2m} \bm{\beta} \, P_{in}(\bm{\beta})   e^{\sum_{j=1}^B(e^{i\eta_j}-1)\sum_{\ell\in\mathcal{K}_j}(L\bm{\beta})^*_\ell (L\bm{\beta})_\ell} 
    \\ & = \int d^{2m} \bm{\beta} \, P_{in}(\bm{\beta}) \, e^{\bm{\beta}^\dagger \mathcal{U}^\intercal \bm{\beta}}  \, ,
    \label{characteristic_function_P_classical}
\end{split}
\end{equation} 
where $\mathcal{U} = L^\intercal H L^*$, and $H=\text{diag}(e^{i\theta_1}-1,\dots,e^{i\theta_M}-1)$ with  $\theta_i = \eta_j$ if $i\in\mathcal{K}_j$.
We can now substitute specify the input state, substitute its $P$ function in the previous expression and explicitly compute the characteristic function by integration. 
\subsection{Thermal state input}
The $P$-function of a generic single-mode Gaussian state with zero displacement and covariance matrix ${\sigma}$ reads \cite{nonlinearBS}
\begin{equation}
    P(\beta) = \frac{2}{\pi\sqrt{\det{{\sigma}-\mathbb{I}_2}}} e^{-2(x \,\, y )({\sigma}-\mathbb{I}_2)^{-1}(x \,\, y )^\intercal} \, ,
    \label{def_p_function}
\end{equation}
where $x$ and $y$ denote the real and imaginary parts of $\beta$, respectively.
The conventions used are such that the covariance matrix of a thermal state $\hat{\nu}_{th}(\overline{n})$ reads ${\sigma} = (2\overline{n}+1)\mathbb{I}_2$, where $\overline{n}=\Tr{\hat{\nu}_{th}(k) \hat{n}}$ is the mean number of thermal photons.
The $P$-function of a 
multi-mode thermal state $\hat{\rho}_{in} = \otimes_{i=1}^m \hat{\nu}_{th}(\overline{n}_i)$ is simply the product of the $P$ functions of single mode thermal states and reads 
\begin{equation}
    P_{th}(\bm{\beta}) = \prod_{i=1}^m \frac{1}{\pi\overline{n}_i} e^{-{(\overline{n}_i)^{-1}}\vert\beta_i\vert^2}  = \mathcal{N} e^{-\bm{\beta}^\dagger D \bm{\beta}}  \, ,
\end{equation}
where $D=\text{diag}((\overline{n}_1)^{-1},\dots,(\overline{n}_m)^{-1}$ and the normalizing factor $\mathcal{N}$ is given by 
\begin{equation}
    \mathcal{N} = \prod_{i=1}^m \left[\frac{1}{\pi\overline{n}_i}\right] \, .
\end{equation}
Hence, we can write the characteristic function as the Gaussian integral
\begin{equation}
    X(\bm{\eta}) =  \mathcal{N}  \int d^{2m} \bm{\beta} \, e^{-\bm{\beta}^\dagger (D-\mathcal{U}^\intercal)\bm{\beta}} \, .
\end{equation}
Let us now highlight the real and imaginary part of the complex vector $\bm{\beta}  = \bm{x} + i\bm{y}$, namely 
\begin{equation}
\begin{split}
    X(\bm{\eta})  & =  \mathcal{N}  \int d^{m}\bm{x}\,d^{m}\bm{y} \, e^{-(\bm{x}^\intercal-i\bm{y}^\intercal) (D-\mathcal{U}^\intercal)(\bm{x}+i\bm{y})} 
    \\ & = \mathcal{N} \int d^{2m}\bm{z} \, e^{-\frac{1}{2}\bm{z}^\intercal Q\bm{z}} \, ,
\end{split}
\end{equation}
where $\bm{z}=(\bm{x},\bm{y})$ and $Q$ is a complex symmetric matrix defined as 
\begin{equation}
    Q = 
    \begin{pmatrix}
        2D-\mathcal{U}-\mathcal{U}^\intercal & i(\mathcal{U}-\mathcal{U}^\intercal) \\ i(\mathcal{U}^\intercal-\mathcal{U}) & 2D-\mathcal{U}-\mathcal{U}^\intercal
    \end{pmatrix} \, .
\end{equation}
Standard multi-dimensional Gaussian integration yields the final result
\begin{equation}
    X(\bm{\eta}) =  \mathcal{N}\frac{(2\pi)^m}{\sqrt{\det{Q}}} \, .
\end{equation}
\subsection{Squashed state input}
A squashed state is a $P$-classical Gaussian state that exhibits vacuum fluctuations in one quadrature and higher fluctuations in the conjugate one.
A single mode squashed state can be parametrized as the following squeezed thermal state $\hat{\rho}_{sq} = \hat{S}(r)\hat{\nu}_{th}(\overline{n})\hat{S}^\dagger(r)$ with $\overline{n}=(e^{2r}-1)/2$, its covariance matrix reading $\sigma_{sq} = \text{diag}(e^{4r},1)$ with $r>0$ without loss of generality.
One then usually sets $r$ such that the squashed state's mean photon number matches that of the squeezed vacuum state it approximates. 
Notice how the matrix ${\sigma}_{sq}-\mathbb{I}_2$ that appears in the $P$ function definition Eq.~\eqref{def_p_function} is now singular, leading to a delta-like divergence in its $P$ function, i.e.
\begin{equation}
    P_{sq}({\beta}) = \sqrt{\frac{2}{\pi\lambda}} e^{-\frac{2x^2}{\lambda}} \delta(y) \, ,
\end{equation}
where $\lambda = e^{4r}-1>0$.
\\
Let us now consider a LON fed with $m$ squashed states, i.e. $\hat{\rho}_{in} = \bigotimes_{i=1}^m \hat{S}(r_i)\hat{\nu}_{th}(\overline{n}_i)\hat{S}^\dagger(r_i)$ with $\overline{n}_i=(e^{2r_i}-1)/2$. The $P$-function of this state clearly reads
\begin{equation}
    P_{sq}(\bm{\beta}) 
    = \mathcal{N} e^{-\bm{x}^\intercal D \bm{x}}\delta^{(m)}(\bm{y}) \, ,
\end{equation}
where $\bm{\beta} = \bm{x} + i\bm{y}$, $D=\text{diag}(\frac{2}{\lambda_1},\dots,\frac{2}{\lambda_m})$ and 
\begin{equation}
    \mathcal{N} = \prod_{i=1}^m \left[\sqrt{\frac{2}{\pi \lambda_i}}\right] \, .
\end{equation}
The characteristic function then reads
\begin{equation}
\begin{split}
    X(\bm{\eta}) & = \mathcal{N} \int d^{m}\bm{x}\,d^m\bm{y}\, e^{-\bm{x}^\intercal D \bm{x}}\delta^{(m)}(\bm{y}) e^{(\bm{x}^\intercal-i\bm{y}^\intercal) \mathcal{U}^\intercal(\bm{x}+i\bm{y})}
    \\ & = \mathcal{N} \int d^{m}\bm{x}\, e^{-\bm{x}^\intercal (D-\mathcal{U}^\intercal) \bm{x}} \, . 
\end{split}
\end{equation}
Upon the symmetrization of the matrix $D-\mathcal{U}^\intercal$ we arrive at
\begin{equation}
    X(\bm{\eta}) = \mathcal{N} \int d^{m}\bm{x}\, e^{-\frac{1}{2}\bm{x}^\intercal Q \bm{x}}  = \mathcal{N} \sqrt{\frac{(2\pi)^m}{\det{Q}}} \, ,
\end{equation}
where $Q = 2D-\mathcal{U}-\mathcal{U}^\intercal$.

\end{document}